\documentclass[journal]{IEEEtran}
\usepackage{graphicx}
\usepackage{amsmath}
\usepackage{amssymb}
\newlength{\zeroheight}
\def\figurewidthsize{0.99 \columnwidth}

\begin{document}



\title{
\large \bf Threshold-based monitoring of cascading outages with PMU measurements of area angle}
\author{
Atena Darvishi
\hspace{1cm} Ian Dobson
\thanks{
The authors are with ECpE dept.,
Iowa State University,
Ames IA USA; dobson@iastate.edu.
We gratefully acknowledge support in part from DOE project ``The future grid to enable sustainable energy systems," an initiative of PSERC, NSF grant CPS-1135825, and the access to the WECC data that enabled this research. The analysis and conclusions are strictly those of the authors and not of WECC.
}}



%
\maketitle

\begin{abstract}  
When power grids are  heavily stressed with a bulk power transfer, it is useful to have a fast indication of the increased  stress when multiple line outages occur.  Reducing the bulk power transfer when the 
outages are severe could forestall further cascading of the outages.
We show that synchrophasor measurements of voltage angles at all the 
area tie lines can be used to  indicate the severity of multiple outages.
These synchrophasor measurements are readily combined into an ``area angle" that can quickly track the severity of 
multiple outages after they occur.
We present a procedure to define thresholds for the area angle that relate to the maximum power that can be transferred through the area 
until a line limit is reached.  
Then in real time we would monitor the area angle and compare it to the thresholds when line outages occur  to determine the urgency (or not) of actions to reduce the bulk transfer of power through the area.
The procedure also identifies exceptional cases  in which separate actions to resolve local power distribution problems are needed.
We illustrate the thresholds and monitoring with the area angle across several states of Northwestern USA.
\end{abstract}


\section{Introduction}
With increasing and variable demands placed on the power transmission system, areas of the power grid are often stressed by bulk transfers through the area.
It is important to be able to quickly monitor the additional stress caused by single and multiple line outages so that appropriate remedial actions can be taken.
Especially in the case of multiple outages, a quick response could prevent further cascading and a blackout.
It is well appreciated that major blackouts have occurred partly due to a lack of situational awareness \cite{2003BlackoutReport}.

\looseness=-1
 In general, synchrophasor  technology makes possible fast and accurate  monitoring and control of  power grids \cite{PhadkeThorpbook}. Synchrophasors  are becoming widespread and operation tools using synchrophasors for wide area monitoring can monitor and manage system stresses to maintain reliability \cite{SlavaGM14,Bhargava04, HuangPESGM08,TaylorIEEEP05}.

 Our method focusses on measuring stress across a particular area of the power system using synchrophasor measurements around the border of the area; that is, synchrophasor measurements at all the tie lines of the area. 
 These synchrophasor measurements around the border of the area are combined into a single angle across the area called the area angle.
 The area angle  obeys circuit laws and is derived from circuit theory  in \cite{DobsonvoltPS12,DobsonPESGM10}. In this paper, we will show that  the area angle  tracks bulk stress caused by line outages inside the area.
We consider the bulk stress to be determined by the maximum bulk transfer through the area that satisfies the line limits inside the area.
 
Some previous works on monitoring power system stress with phasor measurements have focused
on the angle difference between two buses.
Simulations of the grid conditions before the August 2003 USA/Canada blackout show that
increasing large angle differences could be a blackout  precursor \cite{Cummingsslides05}.
Simulations of the New England grid show that
angle differences can discriminate alert and emergency states \cite{ManiPSCE09}.
 A large angle difference between two buses does indicate, in a general sense,  
a stressed power system, but it is difficult to interpret changes in the angle difference or set thresholds.

The advantage of combining the synchrophasor measurements around the border of an area into an area angle is that one is then monitoring stress in that particular area. Then the additional stress due to line outages inside the area can be quickly monitored in real time just after the outages occur.
Furthermore, we will show that our formulation in terms of area angle allows an emergency area angle threshold to be determined based on the maximum 
power transfers through the area. If the monitored area angle exceeds the emergency angle threshold,
the area bulk power transfer should be reduced. 

Given suitable synchrophasor measurements available at a control center \cite{TatePS08},
the calculation of area angle is quick and easy so that the computations can be practical for large real systems. 
We note that synchrophasor measurements around the border of an area can be also advantageous for other applications such as combining AC voltage measurements in a transmission corridor to monitor voltage collapse \cite{LinaPESGM14} or locating line outages in the area \cite{SehwailNAPS12} or stress between areas \cite{LopezPESGM12}. Also we used our method for monitoring single outages \cite{DarvishiNAPS13}. In this paper we seek to monitor the bulk stress  for general line outages in the area that include multiple outages.


In somewhat related work by other authors,  static feasibility boundaries such as those associated with transmission line limits can be determined from grid models with power flows based on SCADA and state estimation. For example,  \cite{GanPES03,Makarov09,CapitanescuPSCC02}  compute minimum security margins under operational uncertainty with respect to thermal overloads.  Also \cite{EPRITRACE91} provides a tool for computation of transfer capability margins.
Our work is different since we use synchrophasor measurements to monitor in real time  the stress with regard to bulk power transfer through areas  due to multiple outages inside the area.
Methods based on the state estimator produce a much more detailed view of a representative power system condition over the SCADA sampling period, and require some computation time
for actionable information.
Our method based on synchrophasors is approximate but faster, and  will work under multiple outage conditions in which 
the state estimator may not readily converge.

\section{Monitoring and thresholds based on angles}
\subsection{Simple example}
\begin{figure}[t]
  \begin{center}
 \includegraphics[width=\figurewidthsize]{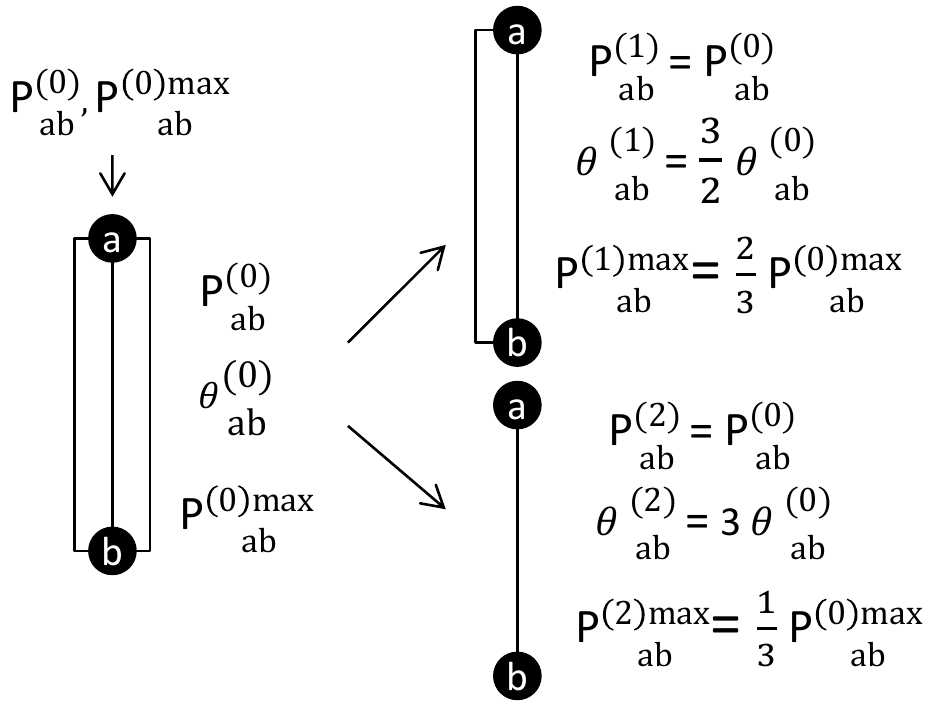}
  \caption{Comparing power and angle in a simple example of 3 parallel lines.}
  \label{SimpleEx3Parallel}
  \end{center}
  \end{figure}

To motivate monitoring with angles, 
 Fig. \ref{SimpleEx3Parallel} shows a simple example of three parallel, lossless lines connecting two buses a and b. 
 DC power flow is assumed and each line has the same power flow limit.
 We  consider three quantities:  $P_{\rm ab}$ is the real power entering bus a and transmitted to bus b,
 $P_{\rm ab}^{\rm max}$ is maximum real power that could enter bus a as determined by the line power flow limits,
 and $\theta_{\rm ab}$ is the voltage angle  between a and b.
 The superscripts (0), (1), and (2) stand for the base, single line outage, and double line outage cases respectively and we consider 
  $P_{\rm ab}$, $P_{\rm ab}^{\rm max}$, and $\theta_{\rm ab}$ in each of these cases.
 
As shown in  Fig. \ref{SimpleEx3Parallel}, the power $P_{\rm ab}$ entering bus a remains the same in all cases, while the angle $\theta_{\rm ab}$ between the buses increases for the single outage and triples in the more severe case of the double outage. The increment in the angle as the outages become more severe is a good indicator of increased stress in the system. That is, the increased stress due to the line outages can be monitored with the angle $\theta_{\rm ab}$, but not with the power $P_{\rm ab}$.

The fact that $\theta_{\rm ab}$ is a good indicator of stress provides the motivation to set a threshold on this quantity. To set a threshold on $\theta_{\rm ab}$ that distinguishes outage severity, 
we first consider the maximum power $P_{\rm ab}^{\rm max}$ entering bus a.
$P_{\rm ab}^{\rm (0)max}$ in the base case is three times the line  limit and, as shown in Fig. \ref{SimpleEx3Parallel},
$P_{\rm ab}^{\rm max}$ decreases as the outages become more severe and the stress  increases.
For example, for a single outage, $P_{\rm ab}^{\rm (1)max}$ is twice the line  limit.
Following the N-1 criteria,
 we may consider that a single outage is the maximum stress level the system can safely tolerate before a 
  line limit is exceeded.
 This stress level corresponds to $P_{\rm ab}^{\rm (1)max}$ in the single contingency case and to the value of $\theta_{\rm ab}$ in the single contingency case when $P_{\rm ab}^{(1)}=P_{\rm ab}^{\rm (1)max}$.
 That is, a threshold $\theta_{\rm ab}^{\rm threshold}$ for $\theta_{\rm ab}$
 is obtained as

  \begin{align}
 \theta_{\rm ab}^{\rm threshold}=\frac{x}{2} P_{\rm ab}^{\rm (1)max}
 \label{thetasimplethreshold}
 \end{align}
 where $x$ is the reactance of one of the lines.
 
 We are interested in how $P_{\rm ab}^{\rm max}$ changes as line outages occur.
$P_{\rm ab}^{\rm max}$ cannot be measured directly (it is calculated by increasing the power  $P_{\rm ab}$ from its current value until a line limit is encountered).
However, we can see from Fig.~\ref{SimpleEx3Parallel} that $P_{\rm ab}^{\rm max}$ is inversely proportional to $\theta_{\rm ab}$, which can be monitored and compared to its threshold $\theta_{\rm ab}^{\rm threshold}$.

If an outage or outages occur, $\theta_{\rm ab}$ increases from its base case value and can be compared to 
the threshold $\theta_{\rm ab}^{\rm threshold}$. $\theta_{\rm ab}\le \theta_{\rm ab}^{\rm threshold}$ indicates 
that line limits are satisfied after the outage(s).  That is, the outage(s) is less severe than the highest-loaded  case of a single outage satisfying the {N-1} criterion, but may well require corrective action to restore 
operating margin.
 On the other hand, 
$\theta_{\rm ab}>\theta_{\rm ab}^{\rm threshold}$ indicates that line limits are violated after the outage(s). These outage(s) are more severe 
 than the highest-loaded  case of a single outage satisfying the {N-1} criterion, and require
 emergency action reducing $P_{\rm ab}$ to resolve the violated line limits.\footnote{If there is a single outage, it is clear from (\ref{thetasimplethreshold}) that  $\theta_{\rm ab}^{(1)}\le\theta_{\rm ab}^{\rm threshold}$ implies line limits satisfied after the single outage and $\theta_{\rm ab}^{(1)}>\theta_{\rm ab}^{\rm threshold}$ implies line limits violated after the single outage.
 If there is a double outage,    
 $
 \theta_{\rm ab}^{(2)}\le\theta_{\rm ab}^{\rm threshold}=
\frac{x}{2} P_{\rm ab}^{\rm (1)max}=x P_{\rm ab}^{\rm (2)max}
$
  implies line limits satisfied after the double outage and $\theta_{\rm ab}^{(2)}>\theta_{\rm ab}^{\rm threshold}$ implies line limits violated after the double outage.}

 The overall strategy is to set thresholds based on the line limits in terms of the economically significant maximum power transfer through the area,
and then convert the threshold on the maximum power transfer to an equivalent threshold on the angle between the buses.
Then monitoring the angle and comparing it to the angle threshold can detect what urgency of  action is needed to reduce the power transfer in order to maintain security.

\subsection{Generalization to an area of a power system}
\label{generalization}

\begin{figure}[t]
  \begin{center}
 \includegraphics[width=3.5in]{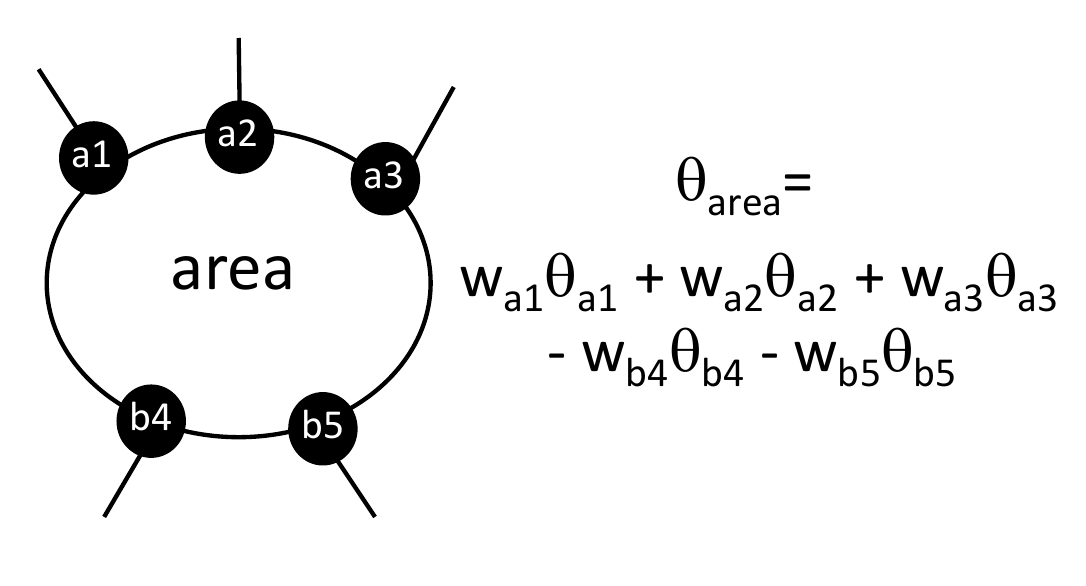}
  \caption{General area of power system  and the area angle.}
  \label{Area}
  \end{center}
  \end{figure}

We generalize the simple example of Fig. \ref{SimpleEx3Parallel}  to the connected area of a power system of Fig. \ref{Area}.
The area is primarily transferring power from the buses in the north (marked a) to the buses in the south (marked b).
The a and b buses together form a complete border of the area, so that removing the a and b buses would entirely disconnect 
the area from the rest of the power system.
There is major generation north of the area and major load south of the area and we are interested in monitoring the area stress 
due to line outages inside the area.

The real power entering the area from the north is the sum of the powers entering the a buses along the tie lines connected to  the a border buses. The  angle difference $\theta_{\rm ab}$ between bus a and bus b in the simple example of Fig. \ref{SimpleEx3Parallel} generalizes to an angle $\theta_{\rm area}$  across the area from  the a buses to  the b buses. The area angle is a new concept derived from circuit theory principles in \cite{DobsonvoltPS12}.
Number the area border buses 1,2,...,$m$  and let the angles at these buses be 
$\theta_{\rm 1}$, $\theta_{\rm 2}$, ... , $\theta_{\rm m}$.
Then the area angle $\theta_{\rm area}$  is computed as a weighted combination of the angles at border buses:
\begin{align}
\theta_{\rm area}=\sum_{j=1}^{m}w_{j}  \theta_{j}
\label{thetanewC}
\end{align}
As might be expected for an angle across an area, the weights on the a buses are generally positive and the weights on the b buses are generally negative.
 The angles at all the a and b buses are obtained from the filtered quasi-steady synchrophasor measurements described in \cite{TatePS08}
 that indicate the settled steady state measurement after the outage.
 
According to \cite{DobsonvoltPS12}, the weights $w$ are computed from the formula
\begin{align}
w=(w_{1}, w_{2}, ... , w_{m})=\frac{\sigma_{\rm a}  B_{\rm eq}}{b_{\rm area}}
\label{w}
\end{align}
  Here $\sigma_{\rm a}$ is the row vector of length $m$ with ones at the positions of the a buses and zeros at the positions of the b buses.
   $B_{\rm eq}$ is the equivalent susceptance matrix of the border buses, which is calculated as the Kron reduction of the 
   area susceptance matrix to the border buses. $b_{\rm area}=\sigma_{\rm a}B_{eq}\sigma_{\rm a}^T$ is the bulk susceptance of the area.
   Overall, it can be seen that the weights $w$  can be obtained from the base case area topology and a DC load flow model of the area. 
   A recent base case of the DC load flow model is generally available \cite{TatePS08}.
   An important detail is that we use the base case DC load flow to compute  the  weights $w$, and do 
   not  attempt to immediately update the DC load flow model based on the outage we are trying to monitor 
   \cite{DobsonIREP10}.

We are interested to obtain area angle thresholds corresponding to specific stress limits in the area and then observe the changes in the area angle caused by different outages inside the area in real time to be notified of different stress condition in the area. To discover the alarm or emergency thresholds based on the area angle, since we quantify stress in terms of the maximum power that could enter the area, we first determine thresholds of the maximum power that could enter the area after the outages and then find out the corresponding area angle thresholds. Then in real time, comparing the area angle after outages with its thresholds notifies us of the different  stress severities of outages inside the area. 

\section{Problem set up}
\subsection{Overview}
\label{Overview}

The first step is to define the  area of interest as explained in section \ref{generalization}, including the particular power transfer through the area, and the border buses of the area at which synchrophasor measurements 
should be made.
The area angle is a weighted combination of these synchrophasor measurements.

There is an offline calculation of actionable thresholds for the area angle and offline identification of any local power redistribution problems that 
are poorly detected by the bulk area angle.
There are a limited number of these local power redistribution problems and they can be separately detected and resolved as explained below.

To apply the area angle online, we monitor the area angle computed from the measurements, and also monitor the local power redistribution problems.
If  there is a change in the area angle and a local power  redistribution problem has not occurred, 
this indicates a change in bulk stress with respect to the transfer through  the area and the line limits.
The area angle after the outage is compared to its precomputed thresholds so that the appropriate action to reduce the power transfer (emergency action, some action needed to restore full security, no action required)
can be chosen.
The emergency threshold distinguishes outages that require emergency action  from outages that require some  action and the alarm threshold
 distinguishes outages that require some action from outages that require no action.

The 
procedures are summarized in the following steps:\newline
1) Offline calculations to set  thresholds and identify local problems
\vspace{-3pt}
  \begin{enumerate}
     \item  For each single outage inside the area, after the outage, calculate  the maximum power that can enter the area before the first line limit is encountered.
     The maximum power that can enter the area for the worst case single outage is the emergency threshold for the maximum power entering the area.
      Also define the alarm  threshold on the maximum power entering the area.       
       \item Set the base case power entering the area to the emergency threshold of the maximum power. Then for all single outages inside the area, calculate the 
      area angle after the outage.
                     \item   By finding outliers to the bulk relationship between the 
                   area angle and the maximum power that can enter the area, outages that cause local power redistribution problems can be identified, and these cases that are poorly detected by the area angle are dealt with separately.
\item               Convert the emergency and alarm thresholds of the maximum power entering the area to the area angle emergency and alarm thresholds
using the bulk relationship between the maximum power that can enter the area and the area angle. 
                                      \end{enumerate}
  
\noindent
2) Online implementation
\vspace{-3pt}
    \begin{enumerate}
       \item In the control center, compute the area angle  from the synchrophasor angles at the border of the area and monitor the occurrence of any of the outages causing  local power redistribution problems.
              \item If outages which are causing local power redistribution problems have not occurred, then compare the area angle to its thresholds to take no action or to take proper action with the appropriate urgency.
              \item If outages that cause local power redistribution problems occur, then take the appropriate local action.
      \end{enumerate}
      
\noindent
We now discuss some of these steps in more detail.

\subsection{Setting thresholds on maximum power and angle}
\label{setlimits}

Since the system is operated with respect to the N-1 criterion for line limits,  no  single line outage will violate a line limit in the base case.
We want to quickly detect from the measurements how the severity of  the outages compares to the worst case single outage. 
To do this, we set the threshold on the maximum power entering the area to be the maximum power entering the area satisfying the line limits when the 
most severe single outage occurs.  Equivalently, this threshold is the minimum of the maximum power entering the area over all single outages, since the most severe single outage restricts the maximum power entering the area the most. 

Now we convert the maximum power emergency threshold into an area angle emergency threshold because we can measure and monitor the area angle. The area angle emergency threshold is the area angle under the worst case single contingency when the power entering the area is equal to the maximum power that could enter the area. 
This area angle threshold is effective because, after the exceptional cases related to local outages are excluded, the area angle approximately increases as the maximum power that could enter the area decreases.
That is, if multiple outages occur and the area angle after the outages is below its emergency threshold, then the corresponding maximum power entering the area
is above its emergency threshold and the outages are comparable in severity to a single outage that does not violate line limits.
After such outages, action may be needed to restore the N-1 security, but no emergency action is required.
On the other hand, if the area angle after the outages exceeds the emergency threshold, then  the corresponding maximum power entering the area
is below its emergency threshold and the outages are comparable in severity to a single outage that violates line limits.
After such outages, emergency action to resolve the problem is appropriate.

It is also useful to set an area angle alarm threshold below which no action is needed.
This alarm threshold corresponds to a suitably small decrease in the maximum power entering the area from the base case maximum power entering the area.
There are many multiple outages that have little effect on the system performance and if the area angle after  these outages is below the alarm threshold, then 
no action needs to be taken.

\looseness=-1
To summarize,  if the area angle after the outage is less than the alarm threshold, the area is safe and we do not need to take any action. If it is between the alarm and the emergency threshold, we need to take some moderate action.  If it is more than the emergency threshold, we need to take emergency action to immediately reduce the bulk power transfer through the area.

\subsection{Finding outages that cause local power redistribution problems}
 It is approximately the case that the area angle gets larger as the maximum power that could enter the area decreases.  
       This relationship describes a bulk property of the area. 
       Plotting this relationship between the maximum power that could enter the area and the  area angle 
       can reveal and identify those exceptional outages that are outliers that do not follow the bulk relationship.\footnote{  
Working with all the single outages in steps 1(a) and 1(b) of section \ref{Overview} identifies all of the outages causing local problems. 
Repeating 1(a) and 1(b)  for a  random selection of double outages can further help to identify these outages. }    
The most common reason for these exceptional line outages causing a local power redistribution problem is proximity to large generation or load inside the area\footnote{
Some grid models combine together lines and generation, especially at  lower voltages, leading to lack of coordination 
between line limits and between line limits and generation; these reduced models can contribute to the exceptional cases.
Also, the area can sometimes be adjusted to exclude large generators or loads.}
\cite{DarvishiNAPS13}.

   
   The exceptional outages and consequent potential overloads are handled separately.
   For example, if the outage is near a larger load inside the area, then we may redispatch the local generation to serve that load. 
   The mitigation or correction of these exceptional outages can be local or by a more wide area scheme, and can use SCADA or synchrophasor data, but in any case, a signal is sent to the control  center when
   one of the exceptional  outages occurs. Our experience so far is that there are a limited number of these exceptional outages to resolve.

\begin{figure}[h]
    \begin{center}
    \includegraphics[width=\columnwidth]{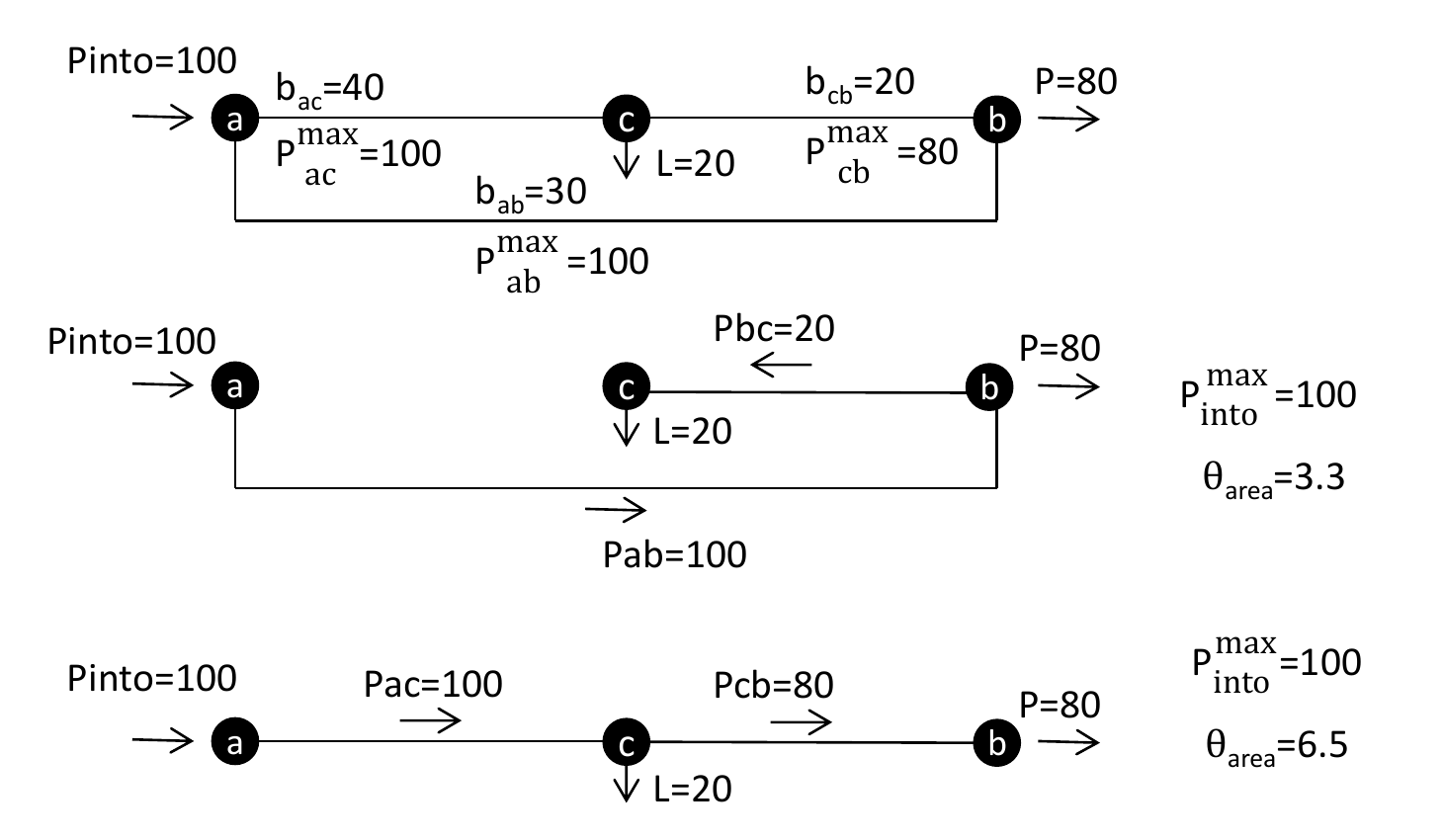}
    \caption{Simple example of a local power redistribution problem}
    \label{SimpleEx2Outage}
    \end{center}
    \end{figure}

 
 We illustrate the effect of large load or generation inside the area in a simple example. Fig.~\ref{SimpleEx2Outage} first shows the base case of a three bus system with buses a and b as border buses and load bus c inside the area,  and then shows the effects of the outage of line a-c and the outage of line a-b. The line limits are chosen to satisfy the N-1 criterion in the base case and are specified in Fig.~\ref{SimpleEx2Outage}.
   We calculate the area angle $\theta_{\rm area}=\theta_{\rm a}-\theta_{\rm b}$ and the maximum power $P_{\rm into}^{\rm max}$ that could enter bus a after each outage  based on the line limits. Fig.~\ref{SimpleEx2Outage} shows that after the outages of line a-c and of line a-b, while the maximum power $P_{\rm into}^{\rm max}$ that could enter bus a is the same, the area angle $\theta_{\rm area}$ after these outages varies considerably.
   The outage of line a-c is an exceptional outage in which load in bus c  makes $\theta_{\rm area}$ less than expected; some power will be redistributed from bus b to bus c in the opposite direction of the bulk power transfer between bus a and bus b. 
 
There is a tradeoff between how closely the area angle tracks the outage severity and the number of exceptional line outages requiring special treatment.  The outage severity tracking can be relaxed by only requiring the correct classification of outages by the thresholds.  In this case the number of exceptional line outages will be smaller and applying the monitoring will be easier. 
A further reduction in the number of exceptional line outages can be achieved if one requires  classification of 
outages only with respect to the emergency threshold. 
  
  Also it should be noted that once an outage of one of the lines which cause local problems is detected, we can have a estimate of the network situation, since we have  observed that  they cause a similar effect when they combine with other outages as when they occur singly, so studying the effect of the exceptional line outages in the single outage case will be very useful in the case of their combination with other outages.    
  
\subsection{Detail of formulation and calculations}
This subsection gives the details of the formulation and calculations of the maximum power that could enter the area and the area angle.
     We use the following notation:   
    \vspace{.1 cm}~
 ~ \newline
   \begin{tabular}{ @{}ll @{}}
  $X$&generic variable\\
  $X^{(i)}$&$X$ evaluated for contingency number $i$. \\
  $X^{k\rm max}$&$X$ evaluated at the maximum power injection \\&case obtained by applying power injection until \\& line $k$ reaches its maximum power flow rating.\\
  $X^{(i){\rm max}}$&$X$ evaluated for the maximum power injection case \\&obtained under  contingency number $i$  \\
  $X^{\rm limit}$& operating limit established for $X$ \\
    \end{tabular} \vspace{1pt}

 To evaluate the maximum power that can enter the area it is necessary to stress the 
 area with additional power injections. These additional power injections are made at each border 
 bus in proportion to the tie line flows entering or leaving that bus, as described later in this subsection.
 To calculate the effect of these power injections, we start with the simpler case of injecting 
 additional power at a particular northern border bus $r$ and removing the same amount of additional power 
 from a southern border bus $s$.
 
To calculate the maximum possible extra power injection at the border buses that satisfies the line limit of all lines after each contingency, we first calculate $\Delta P^{rs(i)k\rm max}$, the maximum possible extra power injection at the border buses satisfying only the line limit of line $k$:
\begin{align}
\Delta P^{rs(i)k\rm max}=\frac{\Delta P^{\text{limit}(i)}_{k}}{\rho^{rs(i)}_k}.
\label{stresskbase}
\end{align}	
$\Delta P^{\text{limit}(i)}_{k}$ is the margin in line $k$ after contingency $i$ which is the power in line $k$ after the contingency $i$ subtracted from the line power limit of line $k$. $\rho^{rs(i)}_k$ is the generation shift factor of line $k$ with respect to the power injection in border buses $r$ and $s$, which can be calculated as 
\begin{align}
\rho^{rs(i)}_k
&=
b_k(e_u^T-e_v^T)(B^{(i)})^{-1}(e_r-e_s).
\label{genshiftfactor}
\end{align}
Here $B^{(i)}$ is the susceptance matrix when line $i$ is outaged, 
$e_r$ is the vector with 1 at entry $r$ and all other entries zero, $b_k$ is the 
susceptance of line $k$, and $u$ and $v$ are the sending and receiving buses of line $k$.

 Then $\Delta P^{rs(i)}_{\rm inj}$, the maximum possible extra power injection at the buses $r$ and $s$ which satisfies all the line limits, is the minimum value of all the injections corresponding to each of the $n$  lines inside the area:
\begin{align}
\Delta P^{rs(i)}_{\rm inj}={\rm Min}\{\Delta P^{rs(i)1\rm max},
\ldots,\Delta P^{rs(i)n\rm max}\}.
\label{stressbase}
\end{align}
We add the extra injection in bus $r$ from (\ref{stressbase}) to  the base case power $P_{{\rm into}r}$ entering bus $r$ to calculate the maximum power $P_{{\rm into}r}^{(i)\rm max}$ that could enter bus $r$ after contingency $i$:
\begin{align}
P_{{\rm into}r}^{(i)\rm max}&=P_{{\rm into}r}+\Delta P^{rs(i)}_{\rm inj}.
\label{maxpintobase}
\end{align}

We expand this calculation from pair $r$ and $s$ to all the border buses of sets $a$ and $b$ and calculate the maximum power that could enter the area through the border buses $a$. In that case the first term on the right hand side of (\ref{maxpintobase}) will change to the power entering the area, which is the sum of the powers entering border buses $a$. To find the second term that is the extra injection considering buses $a$ and $b$ as border buses, we first calculate the generation shift factor of line $k$ with respect to sets $a$ and $b$ and then update (\ref{stresskbase}) and (\ref{stressbase}) accordingly.

We calculate the generation shift factor of line $k$ with respect to sets $a$ and $b$ in the following way. The change in power flow of line $k$ caused by proportional increases in injection in border buses $a$ and $b$ is
\begin{align}
\rho^{ab(i)}_k
&=
b_k(e_u^T-e_v^T)(B^{(i)})^{-1}(e_a-e_b).
\label{genshiftfactorGeneral}
\end{align}
Here $e_a$ and $e_b$ have the entry $\alpha_j$  in positions corresponding to the sets $a$ and $b$ and the rest of the entries zero. The ratio $\alpha_j=P_{{\rm into}j}/P_{{\rm into}a}$ is obtained by dividing the power entering or leaving  border bus $j$ by the total power entering or leaving all border buses in $a$ or in $b$.

To calculate the area angle $\theta_{\rm area}^{(i)}$ corresponding to the  maximum power entering the area with the worst case outage number $i$, 
the system is placed in the condition of limit of the maximum power entering the area with outage $i$, line $i$ is outaged, and then the area angle is evaluated using (\ref{thetanewC}) so that
\begin{align}
\theta_{\rm area}^{(i)}=\sum_{j=1}^mw_j\theta_j^{(i)}.
\label{pmuareaanglebase}
\end{align}

\section{Case study}
\begin{figure}[h]
  \begin{center}
  \includegraphics[width=\columnwidth]{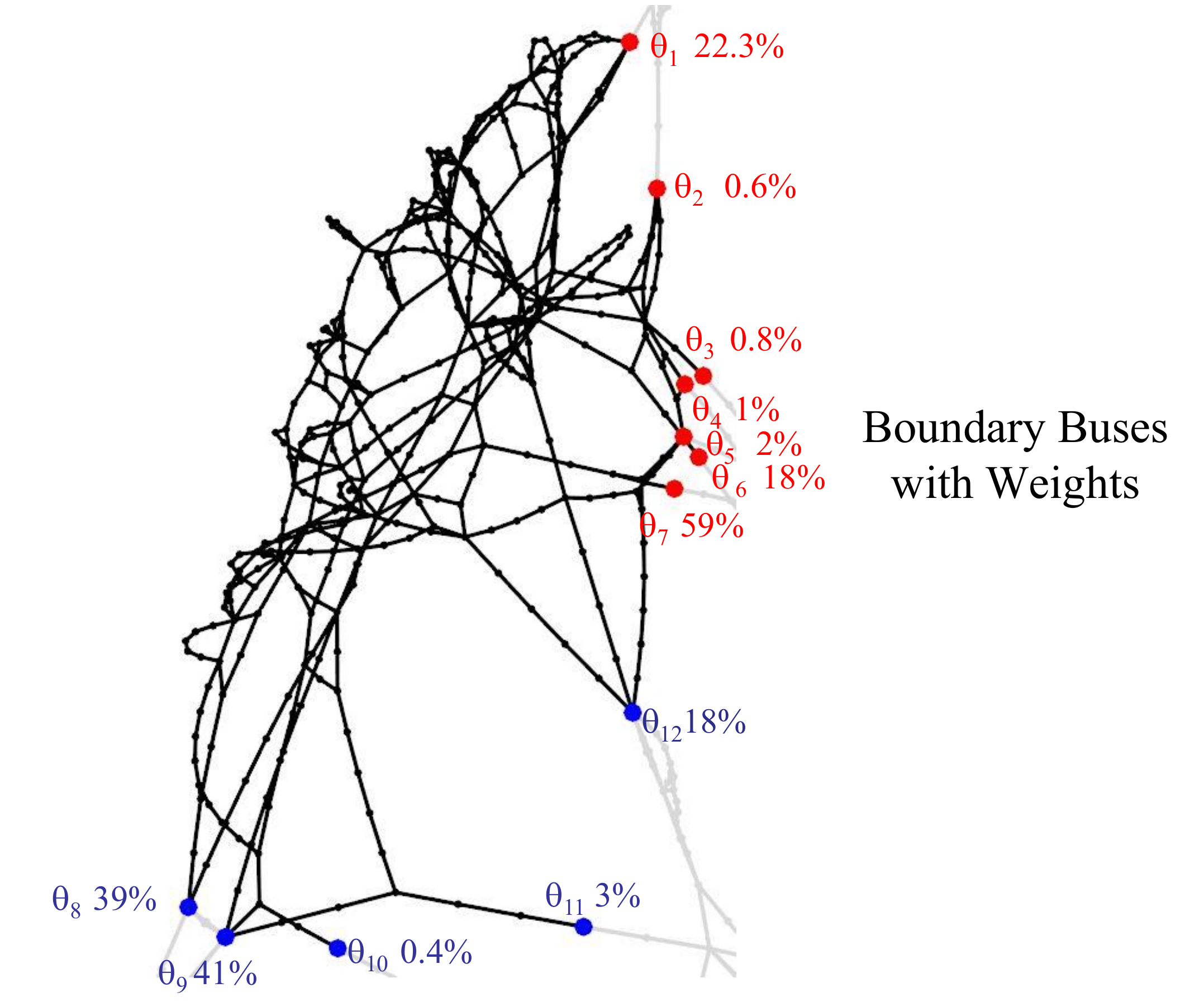}
  \caption{Area of WECC system with area lines in black, north border buses in red and south border buses in blue. Layout is not geographic.}
  \label{pic4ChangeWeccIDGeneral}
  \end{center}
  \end{figure}

We use an area of a 1553 bus reduced model of WECC shown in Fig.~\ref{pic4ChangeWeccIDGeneral}  that covers roughly Washington and Oregon states. The 7 north (and east) border buses are near the borders of Canada-Washington, Washington-Montana, and Oregon-Idaho, and the 5 south border buses are near the Oregon-California border. There are 407 buses and 515 lines inside this area. The bulk power transfer of interest is north to south.

       The  area angle is the following weighted combination of the border bus angles:\footnote{The angle at border bus 5 has a negative weight due to 
       incident lines with 
       negative susceptances arising from grid model reduction.  In practice the measurements with very small weights could be omitted.}
    \begin{align}
\theta_{\rm area} &=\,  0.223\,  \theta_1 + 0.006\,  \theta_2\notag\\& + 0.008 \,\theta_3 + 0.01 \,\theta_4
   - 0.02 \,\theta_5 + 0.18 \,\theta_6+ 0.59 \,\theta_7\notag\\& - 0.39 \,\theta_8 
      - 0.41 \,\theta_9 - 0.004 \,\theta_{10}- 0.03 \,\theta_{11} - 0.18 \,\theta_{12}\notag
  \end{align}

In the following subsections, we do the offline calculation for this area to find area angle thresholds and the lines which cause local power redistribution problems. 
Then, to test how it would perform when monitoring multiple outages online, we use a random sample of triple outages to check that the  thresholds can discriminate outage severity. 

\subsection{Offline calculation to find thresholds and local power redistribution problems}
\begin{figure}[h]
\begin{center} \includegraphics[width=\figurewidthsize]{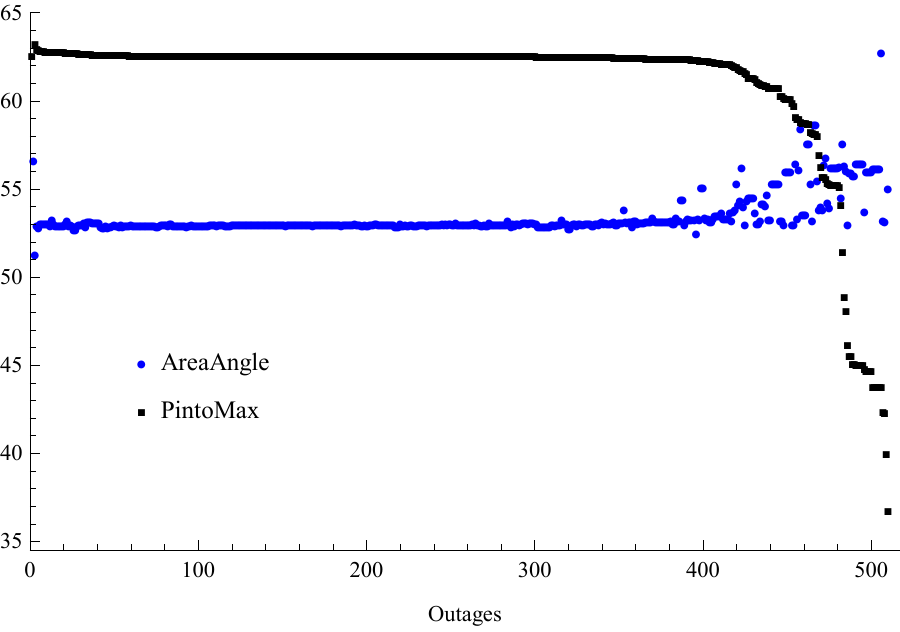}
\caption{Area angle $\theta_{\rm area}^{(i)}$ in degrees, and maximum power into the area in per unit for  all single outages inside the area.
Horizontal axis is outage number.}
\label{pl4WeccIDGeneral3}
\end{center}
\end{figure}

\paragraph{Finding emergency and alarm thresholds of maximum power transfer}
We take out all single lines inside the area in turn and calculate the maximum power that could enter the area for each. It should be noted that the maximum power that could enter the area is related to the line limits and does not depend on the base case load level. We order the outages by decreasing amount of the maximum power transfer so that the outages are ordered with increasing severity and show the results in Fig.~\ref{pl4WeccIDGeneral3}. Then it can be seen that the lowest value of the maximum power transfer (the worst case single outage) is near 35 pu and so we initially set the emergency maximum power threshold to be 35 pu. 
 However, in the case considered, the worst case single outage 
 turns out to be an exceptional outage in step (c) below. Thus there arises a choice, after step (c)  in the detail of setting the emergency threshold 
 of whether the worst case outage should be the worst case outage over all outages or 
 the worst case over the non-exceptional outages. 
 Since we are looking for the thresholds with respect to bulk power transfer, after step (c), we decide to set the emergency threshold according to the worst case non-exceptional outage 
 and revise the emergency threshold for the maximum power entering the threshold accordingly  to 40 pu.
Considering that the  maximum power transfer for the base case (no outage) is 62.5 pu, we set the alarm threshold on the 
maximum power transfer to 60 pu.
(The maximum power transfer decreases more quickly below 60 pu so that those outages start to be more severe.)

\paragraph{Calculate the area angle for outages}

\begin{figure}[h]
\begin{center}
\includegraphics[width=\figurewidthsize]{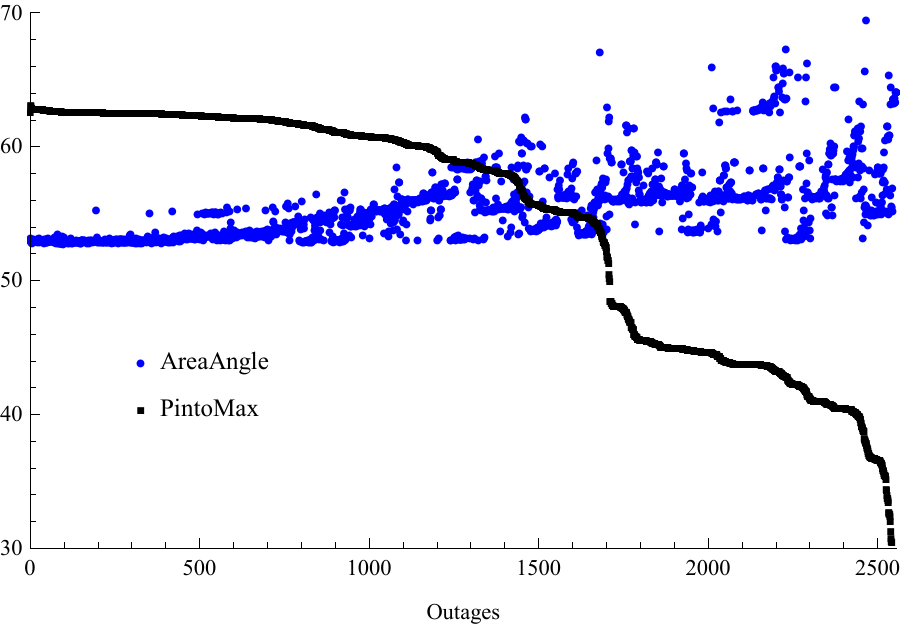}
\caption{Area angle $\theta_{\rm area}^{(i)}$ in degrees, and maximum power into the area in per unit for a random sample of double outages inside the area.
}
\label{Pl472SevereRBbWeccCascade2}
\end{center}
\end{figure}

This calculation is done  with the bulk power transfer set to 35 pu (maximum power transfer for the worse single outage).
We find the area angle after all single outage, and then order the outages according to  increasing severity. We also do the same calculation for a random combination of double outages, since this can help us to find the local problems more easily. Figs.~\ref{pl4WeccIDGeneral3} and  \ref{Pl472SevereRBbWeccCascade2} show the results ordered by outage severity for single and double outages.

\begin{figure}[h]
    \begin{center}
    \includegraphics[width=\figurewidthsize]{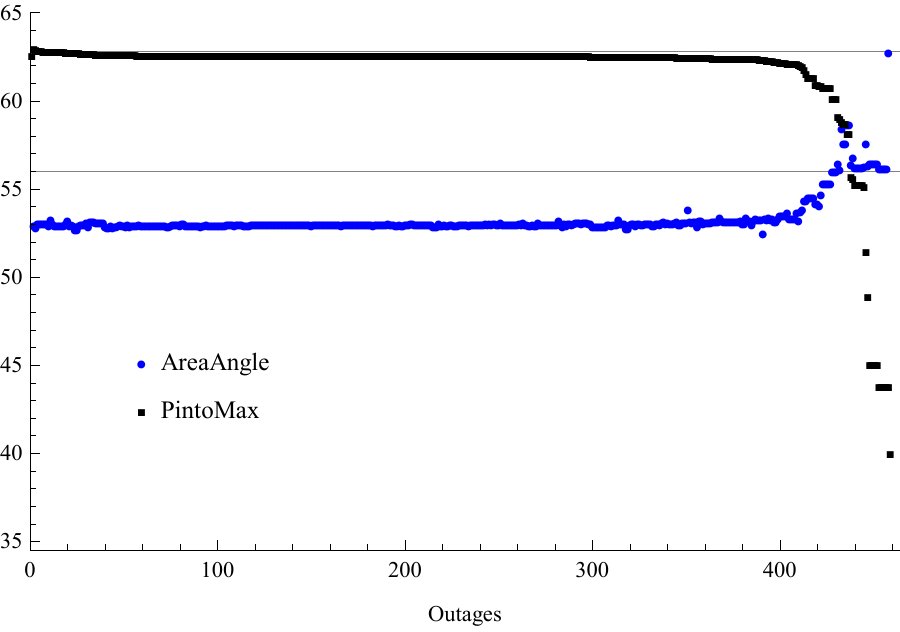}
    \caption{Area angle $\theta_{\rm area}^{(i)}$ in degrees, and maximum power into the area in per unit for all non-exceptional single outages inside the area.  Horizontal line is area angle alarm threshold.
    Horizontal axis is outage number.}
    \label{pl4RefinedWeccIDGeneral3}
    \end{center}
    \end{figure}

\begin{figure}[h]
  \begin{center}
  \includegraphics[width=\figurewidthsize]{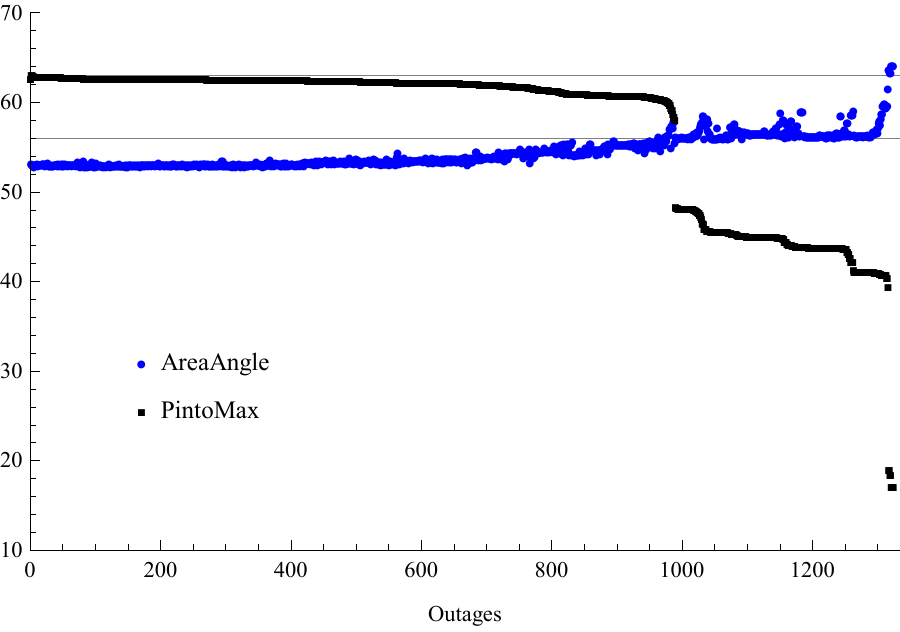}
  \caption{Area angle $\theta_{\rm area}^{(i)}$ in degrees, and maximum power into the area in per unit for a random sample of non-exceptional double outages in the area.
  Upper horizontal line is area angle emergency threshold and lower horizontal line is area angle alarm threshold.
  Horizontal axis is outage number.}
  \label{Pl452RRefined72SevereRBbWeccCascade2}
  \end{center}
  \end{figure}

\paragraph{Finding the exceptional outages}
 As we can see in Figs.~\ref{pl4WeccIDGeneral3} and \ref{Pl472SevereRBbWeccCascade2}, generally the area angle increases as the maximum power that could enter the area decreases. But there are some outliers that indicate exceptional outages that do not follow this pattern whose severity is  poorly indicated by the area angle. These exceptional outages are associated with local power redistribution problems and appear in the form of individual points in the single outage case and as sets of points in the double case (in the double outage case, the combination of each exceptional line outage with all the other line outages makes a set of points). In the case studied, there are about 54 outlier lines out of 515 lines inside the area. Removing these lines from Fig.~\ref{pl4WeccIDGeneral3} yields Fig.~\ref{pl4RefinedWeccIDGeneral3}.
 Only about 30 of these 54 outliers are  of concern in causing a local power redistribution problem. Removing the 30 main outliers from Fig. \ref{Pl472SevereRBbWeccCascade2} yields  Fig.\ref{Pl452RRefined72SevereRBbWeccCascade2}.

\paragraph{Convert the thresholds of the maximum power entering the area to angle thresholds}

We use Figs.~\ref{pl4RefinedWeccIDGeneral3} and \ref{Pl452RRefined72SevereRBbWeccCascade2} to convert the maximum power emergency threshold of 40 pu to the emergency area angle threshold of 63 degrees.
For all the severe double outages in Fig.~\ref{Pl452RRefined72SevereRBbWeccCascade2} that reduce the maximum power that could enter the area below 40 pu, the area angle is above 63 degrees.

We can use either Fig.~\ref{pl4RefinedWeccIDGeneral3} or Fig.~\ref{Pl452RRefined72SevereRBbWeccCascade2} to convert the alarm power threshold of 60 pu to the area angle threshold of 56 degrees.
For all the moderate severe outages that reduce the maximum power transfer to between 40 and 60 pu, the area angle is between 56 and 63 degrees.

\subsection{Test for online implementation}

For online monitoring, we would first check using SCADA or synchrophasor data whether the line outage is one predetermined to cause a local problem. If the outage is an outage causing local problems, this is resolved by local actions. If the outage  is not an outage causing local problems, we compute the area angle from the synchrophasor measurements of angles at the border buses after the outage and
compare this  area angle to the area angle thresholds to determine if the outage is safe, moderately severe, or severe.

\begin{figure}[h]
\begin{center}
\includegraphics[width=\figurewidthsize]{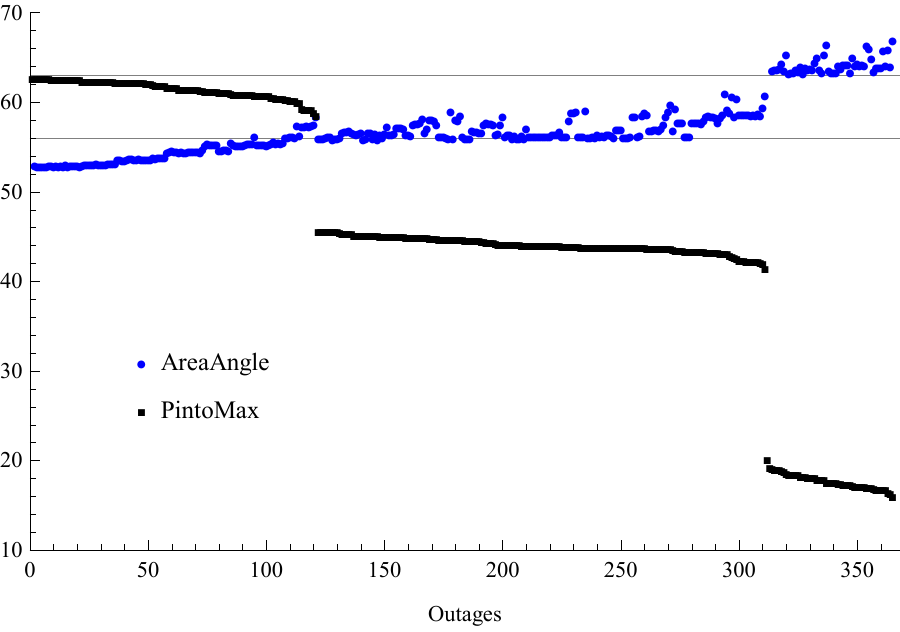}
\caption{Area angle $\theta_{\rm area}^{(i)}$ in degrees, and maximum power into the area in per unit for a random sample of non-exceptional triple outages in the area.
}
\label{Pl414Refined22SevereRBbWeccCascade3}
\end{center}
\end{figure}

To test the method, we randomly sampled triple outages from all lines except the 30 lines that cause local problems and computed the area angle and maximum power entering the area after each of these triples outages. We ordered the results and plotted them in Fig.~\ref{Pl414Refined22SevereRBbWeccCascade3}. 
As can be seen in Fig.~\ref{Pl414Refined22SevereRBbWeccCascade3}, 
for the most severe triple outages (numbered from 310 to 350) that reduce the maximum power coming to the area below 40 pu, the area angle is above the emergency threshold of 63 degrees and for all triple outages numbered from 100 to 310 which decrease the maximum power transfer from 60 pu  to 40 pu, the area angle is between 56 and 63 degrees, and for the rest of triple outages numbered less than 100 which decrease the maximum power transfer only to 60 pu, the area angle is below 56 degrees. 

 The emergency threshold is also effective for multiple outages in discriminating emergency cases in which line overloads are caused, since for all the multiple outages that are above the emergency limit of 63 degrees the maximum power that could enter to the area is below 35 pu,  the maximum power transfer for the worst case single outage over all single outages. This implies that for all multiple outages above the emergency limit  some line limits are violated.

\subsection{Trade off between classification accuracy and the number of exceptional outages}
Fig.~\ref{Pl414Refined22SevereRBbWeccCascade3} shows that area angle can track the severity and can classify the outages into alarm and emergency cases with 30 exceptional line outages. But if one is less interested in tracking severity and only interested in discriminating the emergency outages, the number of exceptional outages can be reduced to 15. 
\begin{figure}[h]
\begin{center}
\includegraphics[width=\figurewidthsize]{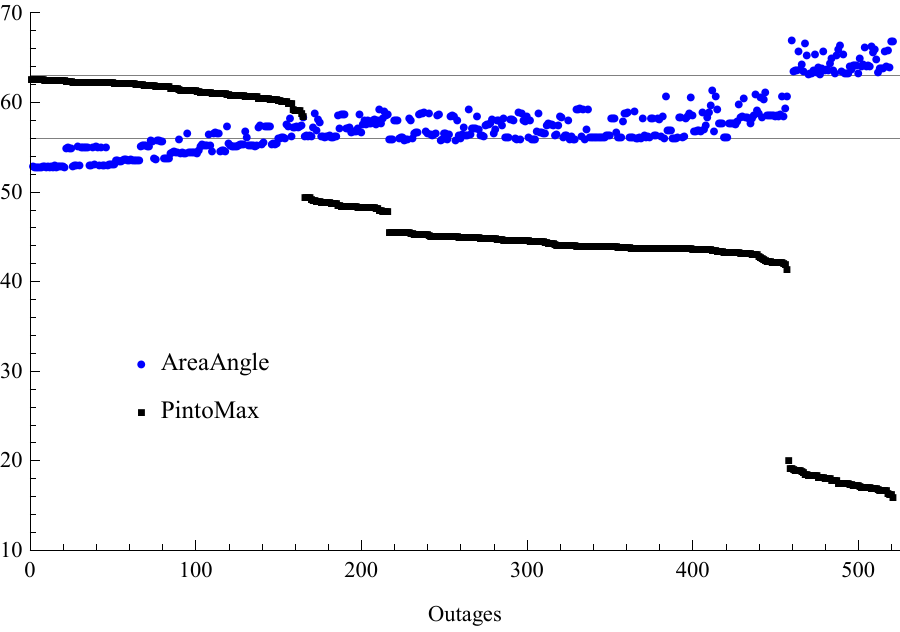}
\caption{Area angle $\theta_{\rm area}^{(i)}$ in degrees, and maximum power into the area in per unit for a random sample of triple outages in the area with 15 exceptional outages excluded.
Horizontal axis is outage number.}
\label{Pl416Refined22SevereRBbWeccCascade3}
\end{center}
\end{figure} 
Fig.~\ref{Pl416Refined22SevereRBbWeccCascade3} shows this more relaxed classification for triple outages. Fig.~\ref{Pl416Refined22SevereRBbWeccCascade3}  preserves the emergency threshold, but loses the exact track of the severity by the area angle and the exact classification between the safe and moderate outages.

We need to monitor the 15 exceptional outages and resolve them separately with local actions. 
Our experience is that these outages cause the same discrepancy in area angle (usually underestimating, but a few overestimating severity)
in both the single and double outage cases.
Therefore appropriate actions can be deduced from the single action case.

\section{conclusion}
   
   We combine a limited number of synchrophasor measurements at the border of a suitable chosen area of the power system  to 
   obtain an area angle that can quickly monitor the severity of multiple outages inside the area. 
   This capability could help to mitigate cascading outages in the early, slower stages of  cascading.
   Our approach relates the area angle to the maximum power that can be transferred through the area for a particular   bulk power transfer direction 
   through the area. 
  
   We describe a procedure to set a meaningful emergency threshold for the area angle after multiple outages according to the 
   worst case single outage. This worst case corresponds to the N-1 criterion.
   Moreover, our results show that this emergency threshold is also effective for multiple outages in discriminating emergency cases in which line overloads are caused. The procedure also identifies line outages associated with local power transfer problems that can also limit the bulk power transfer; these 
   line outages are addressed separately by separate monitoring and control actions.
      
  The angle severity and thresholds are obtained by considering the bulk power transfer of power throughout the area as limited by overloads of lines inside the area.
   This formulation limits the monitoring to the limits on this bulk transfer, but  has 
   the significant benefit that if the area angle exceeds the threshold, then the mitigating action of reducing the transfer is clear.
   This approach yields actionable information based on synchrophasors including thresholds and actions rather than 
vaguely   indicating   there is a problem without  remedies to resolve it.

\end{document}